\documentclass[twocolumn, 11pt]{article}
\usepackage{array, float}
\usepackage[left=18mm, right=18mm, top=26mm, bottom=24mm]{geometry}
\usepackage{newtxtext}
\usepackage{amsmath, amssymb}
\usepackage{mathrsfs}
\usepackage{tabularx}
\usepackage{graphicx}
\usepackage{dcolumn}
\usepackage{setspace}
\usepackage{pgfpages}
\makeatletter
\usepackage{dcolumn}
\usepackage{url}
\usepackage{comment}
\usepackage{lineno}
\usepackage{cite}
\usepackage[numbers,sort&compress]{natbib}

\usepackage{abstract}

\usepackage{setspace}
\usepackage[font=small, format=plain, labelfont=bf, up, textfont=normal, up, justification=justified,singlelinecheck=false]{caption}

\usepackage[affil-it]{authblk}

\newcommand{\win}{w_{\mathrm{in}}}
\newcommand{\wout}{w_{\mathrm{out}}}

\let\OLDthebibliography\thebibliography
\renewcommand\thebibliography[1]{
\OLDthebibliography{#1}
\setlength{\parskip}{0pt}
\setlength{\itemsep}{0pt plus 0.3ex}
}

\title{Partner selection and evolution of out-group avoidance}
\author{Hirofumi Takesue\thanks{Electronic address: \texttt{hir.takesue@gmail.com}}}
\affil{Faculty of Law and Politics, Tokyo Metropolitan University}
\date{}

\begin{document}

\twocolumn[

\maketitle

\begin{onecolabstract}
The preferential treatment of in-group members is widely observed. This study examines this phenomenon in the domain of cooperation in social dilemmas using evolutionary agent-based models that consider the role of partner selection. The model considers a repeated prisoner's dilemma game, in which agents belong to one of two groups that are distinguished by the continuation probability of pair interactions. On the basis of the behavior in the last round and the group affiliation of the partner, each individual selects to cooperate, defect, or stop interactions and search for a different partner. The results of simulation demonstrated that agents adopt cooperative strategies, including tit-for-tat and out-for-tat, toward in-group members. By contrast, agents stop interactions immediately after pair formation without observing their partner's behavior when they are paired with an out-group individual. Higher continuation probability with in-group partners hinders interaction with out-group individuals. Our results imply the importance of avoidance in intergroup interactions in social dilemmas and might explain in-group favoritism without enmity toward out-groups. 
\\\\
\end{onecolabstract}
]
\saythanks

\section*{Introduction}
In-group favoritism is a robustly observed human tendency. Despite the critical importance and pervasiveness of cooperation in human societies \citep{Rand2013a}, cooperation is often conditional on group affiliation. People distinguish between in- and out-groups and cooperate at different levels with them \citep{Balliet2014}. Preferential treatments of in-group members have been observed in artificial groups constructed in labs \citep{Tajfel1970, Dunham2018}, as well as in natural groups existing in societies, such as ethnic groups \citep{Whitt2007}, religious groups \citep{Enos2016}, and political parties \citep{Hahm2024}. In-group favoritism is also observed in other contexts, such as in the positive evaluation of the traits of in-group members \citep{Dunham2018}.

Evolutionary game theory provides insight into the evolution of pervasive in-group favoritism \citep{Masuda2015}. Research has investigated in-group favoritism through the adoption of the models of social dilemmas \citep{Riolo2001}. The models of the evolution of cooperation \citep{Nowak2006, Sigmund2010}, therefore, have contributed to our understanding of the evolution of in-group favoritism. Studies have identified the conditions that can produce a higher propensity to cooperate with members of in-groups than of out-groups. Repeated interactions support the evolution of cooperation \citep{Axelrod1981, Rossetti2023}, but this mechanism, namely, direct reciprocity, combined with higher interaction frequencies with in-group members, can lead to the emergence of in-group favoritism \citep{Rand2013a}. Models of indirect reciprocity indicate that reputation supports cooperation \citep{Nowak2005a, Takacs2021, Xia2023}, but rich reputation information on in-group members can support in-group favoritism \citep{Masuda2007a}. Interactions within networks sustain cooperation \citep{Szabo2007, Perc2017}, but this mechanism also supports in-group favoritism when tags are introduced that differentiate groups \citep{Hochberg2003, Kim2014, Jensen2019}. Cooperation can serve as a costly signal for preferred traits \citep{Gintis2001}, but group-dependent costly signaling and in-group favoritism can coevolve \citep{Macanovic2024}. 

This study examines partner selection mechanisms to investigate the emergence of preferential treatment of in-group members. Models in evolutionary games generally assume random matching or interactions on fixed networks and that players cannot choose interaction partners. Partner selection models, by contrast, stress the importance of choosing interaction partners in the evolution of cooperation \citep{Barclay2016}. The literature has explored this idea in various ways. In repeated games, the discontinuance of interactions that are conditional on the defection of the partner has been found to support cooperation \citep{Vanberg1992, Izquierdo2010, Izquierdo2014, Wubs2016}. In models of dynamic networks, agents can adjust links with their neighbors to avoid noncooperative individuals \citep{Zimmermann2004, Pacheco2006, Fu2008, VanSegbroeck2008, Perc2010}. Mobility in networks supports cooperators because cooperators prevent exploitation by walking away from non-cooperators \citep{Aktipis2004, Helbing2009, Cardinot2019}. The framework of partner selection has been applied to studies of various types of prosociality. For instance, partner choice fosters equal split in both dyadic and group bargaining games \citep{Chiang2008, Andre2011, Takesue2017a}, and a combination with ability and effort can lead to the emergence of contribution-based equity \citep{Debove2017, Takesue2017}. Behavioral experiments show that opportunities to select interaction partners actually support prosocial behavior \citep{Chiang2010, Fehl2011, Rand2011, Shirado2013, Debove2015}. Cross-cultural research has shown that the mobility of relationships in society accompanies high trust levels \citep{Thomson2018}. 

This paper introduces a partner selection mechanism by allowing a leave option in the repeated prisoner's dilemma game (PDG) \citep{Vanberg1992, Izquierdo2010, Izquierdo2014, Wubs2016}. Agents participate in repeated games with randomly matched agents, but they can stop interactions before they are begun or after observing the partner's behavior. In addition to partner control exerted through the tit-for-tat strategy \citep{Axelrod1981}, this option permits the out-for-tat strategy \citep{Yamagishi1994}. In out-for-tat, agents respond to cooperation with cooperation but respond to noncooperation with the leave option. In addition, to consider group structure, we assumed that the continuation probability in repeated games is higher for in-group pairs \citep{Rand2013a}. In particular, exogenous shocks that dissolve pair relationships are more likely for pairs consisting of members of different groups. Community structure, i.e., group structure, is widely observed across social networks. Individuals are located within tightly connected groups and links across those groups are scarce \citep{Girvan2002}. This network structure results in frequent and stable relationships among in-group members, and this may support the assumption regarding continuation probability \citep{Masuda2015}. 

The investigation of this model using evolutionary agent-based simulations demonstrated the emergence of out-group avoidance. In the simulations, we allowed agents to adopt different behavioral rules toward in- and out-group partners. When paired with an in-group member, agents adopt conditionally cooperative strategies, such as tit-for-tat and out-for-tat. By contrast, agents do not initiate interactions with an out-group partner. Strategies for out-group interactions tend to select the leave option at the first round of pair interactions, and pairs dissolve before games effectively begin. This out-group avoidance arises even when the value of continuation probability with out-group agents is high; cooperation can evolve when these values are applied to an entire population, without effective group distinction. These outcomes imply that a default strategy in interaction with out-group members can be avoidance. Stable relationships with in-group members assured by group structure can reduce the need to interact with out-group members, which can foster out-group avoidance from the outset. 

\section*{Model}
We consider a population consisting of $N$ agents. Each agent belongs to one of two groups of equal size. Agents participate in repeated PDG with a leave option ($L$) \citep{Szabo2002}, as well as the options cooperation ($C$) and defection ($D$). The payoff for mutual cooperation (defection) is $b-1 (0)$, whereas unilateral cooperation (defection) leads to a payoff of $-1 (b)$. Here, the parameter $b$ represents the benefit of cooperation. If at least one of the two paired agents chooses the leave option, both agents gain a payoff of $\sigma$.

\begin{figure*}[tbp]
\centering
\vspace{5mm}
\includegraphics[width = 110mm, trim= 0 0 0 0]{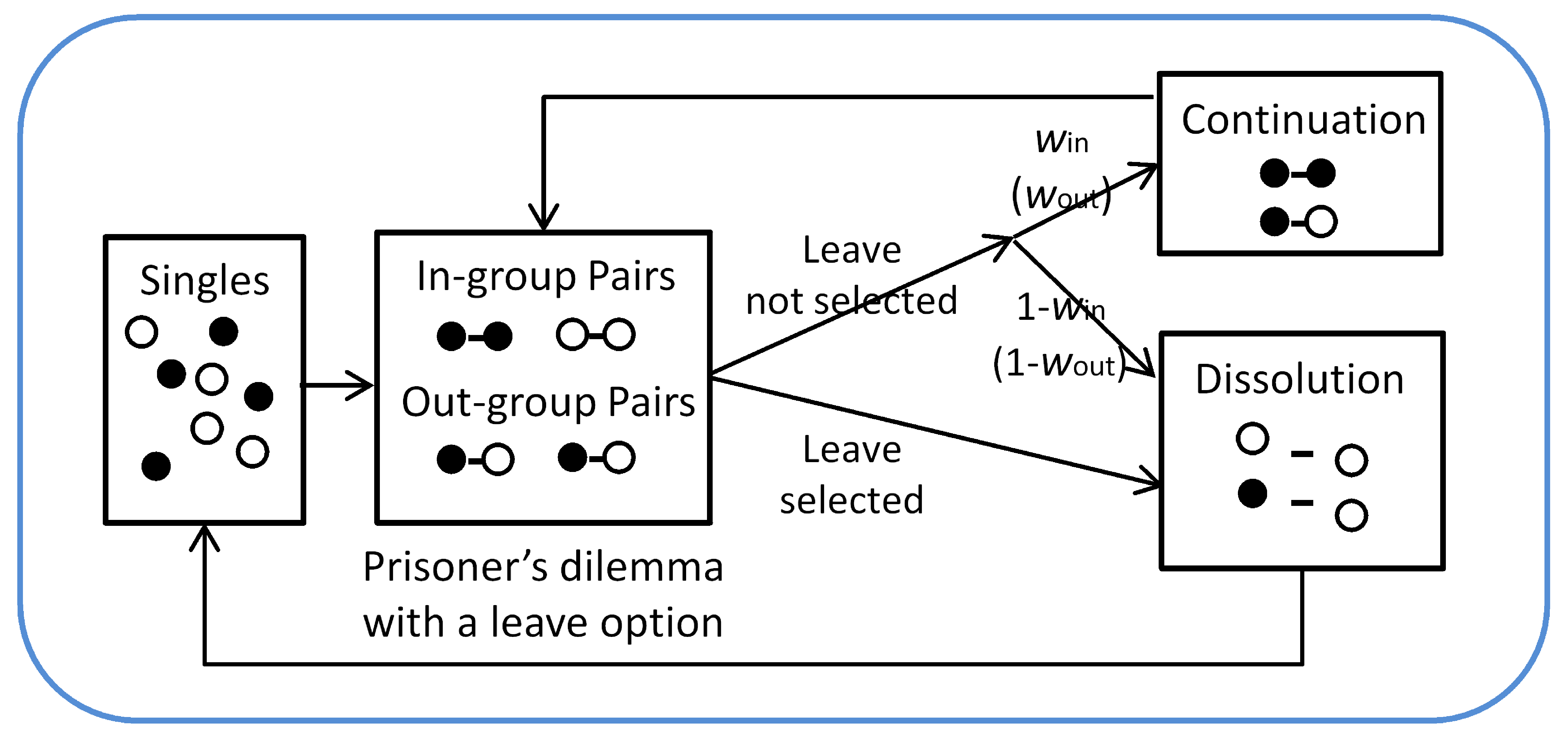}
\caption{\small Schematic explanation of the sequence of events in each period.}
\label{process}
\end{figure*}

The following events repeat for $T$ periods across one generation (Figure~\ref{process} presents the sequence of events in each period in schematic form). (i) Pair formation: pairs are constructed randomly from single agents. The pool of single agents changes dynamically each period. At the outset of each generation, all agents are single. (ii) PDG: paired agents participate in the PDG with the leave option and acquire payoffs. (iii) Endogenous pair dissolution: if at least one paired agent chooses $L$, the pair dissolves, and the two agents return to the pool of single agents. (iv) Exogenous pair dissolution: pairs may dissolute probabilistically. The continuation probability for pairs consisting of in-group members is $\win$, whereas that for pairs consisting of members of different groups is $\wout$. We assume that $\win > \wout$, meaning that exogenous pair dissolution is less likely to occur when agents interact with members of the same group. Through this repeated process, agents accumulate payoffs that affect reproduction. 

We consider strategies to determine actions conditional on the partner's action in a previous round \citep{Izquierdo2010, Wubs2016}. Unlike previous studies, we assume that actions depend on group affiliation as well. Strategies consist of six elements, and each element can take $C$, $D$, or $L$. The first element specifies the action that is selected in the first period when a pair is formed with an in-group member. The second and third elements specify the action after observing an in-group partner selecting $C$ and $D$ in the last round, respectively. In a similar manner, the other three elements specify action toward an out-group partner. For instance, the strategy $CCD\_DDD$, means that agents adopt tit-for-tat toward an in-group partner, whereas they always select $D$ when interacting with an out-group partner. We adopt the letter $X$ to report the summed frequency of multiple strategies. For instance, $CCL\_LXX$ includes all strategies that specify out-for-tat toward an in-group partner and select $L$ in the first round when paired with an out-group member. The last second elements in this example do not have behavioral consequences because pairs dissolve in the first round regardless of the action of the out-group partner. The strategies affect payoffs in the PDG and the occurrence of endogenous pair dissolution, as explained above. 

Reproduction occurs after the PDG is repeated for $T$ rounds. Each agent copies the strategy of an agent selected with probability proportional to transformed payoffs \cite{Lipowski2012}. The transformed payoff for agent $i$ is $\exp (\beta \Pi_i / T)$, where $\Pi_i$ is the payoff accumulated in $T$ rounds, and $\beta$ is selection intensity. In the simulation results reported below, the pool of copied agents is the entire population. However, we confirm that the basic patterns remain the same when the pool of copied agents is restricted to in-group members. Following payoff-based reproduction, mutation can occur. With a probability of $\mu$, each of the six elements may independently takes one of three options: $C$, $D$, or $L$. It is possible that only some elements mutate. 

To enhance statistical accuracy, the simulation first lasts at least $10^4$ generations, following which we recorded the strategy frequencies for at least $10^5$ generations. In addition, we conducted at least five simulations for each combination of parameters, and we calculated the average strategy frequency. In the analysis below, we focused on parameter values that assure cooperation among in-group members. Certain parameter values, such as small $b$ (small benefit of cooperation) and small $\win$ (unstable relationships) have clearly negative impacts on (in-group) cooperation. However, we analyzed the cases in which in-group cooperation evolved and focused on the difference between behaviors toward in- and out-group members. 

\section*{Results}
\begin{figure}[tbp]
\centering
\vspace{5mm}
\includegraphics[width = 85mm, trim= 0 0 0 0]{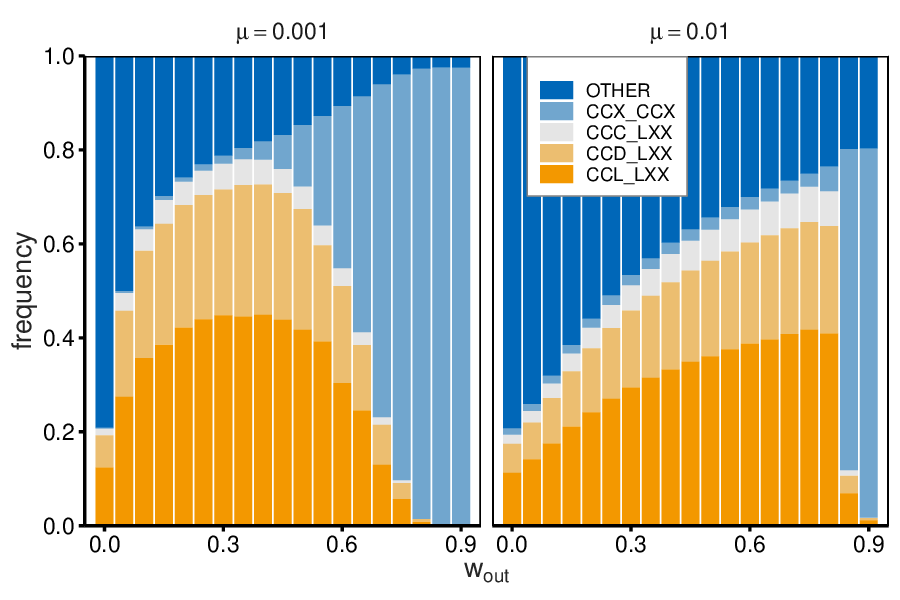}
\caption{\small Frequencies of strategies are reported as a function of $\wout$ when $\win = 0.9$. Agents cooperate with in-group members while avoiding interactions with out-group members from the outset. Indiscriminate cooperation emerges when the values of $\wout$ is sufficiently large. Other parameters: $N = 1000, L = 100, b = 3, \sigma = 0$, and $\beta = 1$.}
\label{w090_sigma000}
\end{figure}

We here report how different values of continuation probability with out-group partners do or do not foster cooperation. We focused on four types of strategies. The first three types ($CCC\_LXX$, $CCD\_LXX$, and $CCL\_LXX$) represent cooperative behavior toward in-group members and avoid interaction with out-group members from the outset. The fourth type ($CCX\_CCX$) specifies cooperation regardless of group affiliation, unless the partner chooses defection. Figure~\ref{w090_sigma000} reports the frequencies of strategies as a function of $\wout$ when $\win = 0.9$. The strategy of selective cooperation with in-group members and out-group avoidance flourishes for small values of $\wout$. Agents frequently adopt tit-for-tat ($CCD$) or out-for-tat ($CCL$) strategies toward in-group partners. By contrast, cooperation that does not discriminate between the two groups evolves as the value of $\wout$ approaches that of $\win$. This transition is steeper where there is a high mutation rate (0.01). Note that out-group avoidance is not prominent for small values of $\wout$ as strategies toward out-group agents have small impacts on payoffs due to frequent exogenous dissolution.  

\begin{figure}[tbp]
\centering
\vspace{5mm}
\includegraphics[width = 85mm, trim= 0 0 0 0]{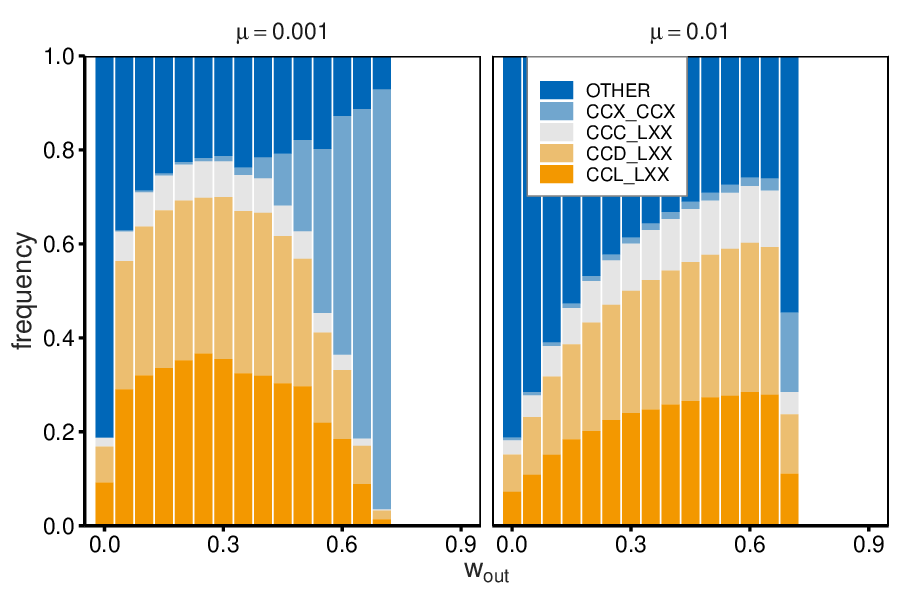}
\caption{\small Frequencies of strategies are reported as a function of $\wout$ when $\win = 0.7$. These results are similar to those obtained when $\win = 0.9$. Other parameters: $N = 1000, L = 100, b = 3, \sigma = 0$, and $\beta = 1$.}
\label{w070_sigma000}
\end{figure}

This pattern can be observed with different values of $\win$. Figure~\ref{w070_sigma000} reports the same quantities when $\win = 0.7$. There is some difference resulting from this setting. For example, a smaller continuation probability ($\win = \wout = 0.7$) is not sufficiently large to allow the evolution of indiscriminate cooperation when $\mu = 0.01$. Here, we still observe in-group favoritism (and out-group favoritism that is included in the category of other strategies and has almost the same frequency as in-group favoritism). Despite this difference, a small continuation probability with out-group members fosters cooperation toward in-group members and the avoidance of out-group members.

To understand the role of different continuation probabilities, we conducted additional simulations wherein the continuation probability is equal for in- and out-group members ($\win = \wout$), which imitates a population without a group structure while preserving basic simulation settings. We compared this simulation with the main simulation results, which assume different continuation probabilities. This simulation is rather artificial because there is no effective distinction between the two groups, but it is useful for understanding the role of different continuation probabilities. 

\begin{figure}[tbp]
\centering
\vspace{5mm}
\includegraphics[width = 85mm, trim= 0 0 0 0]{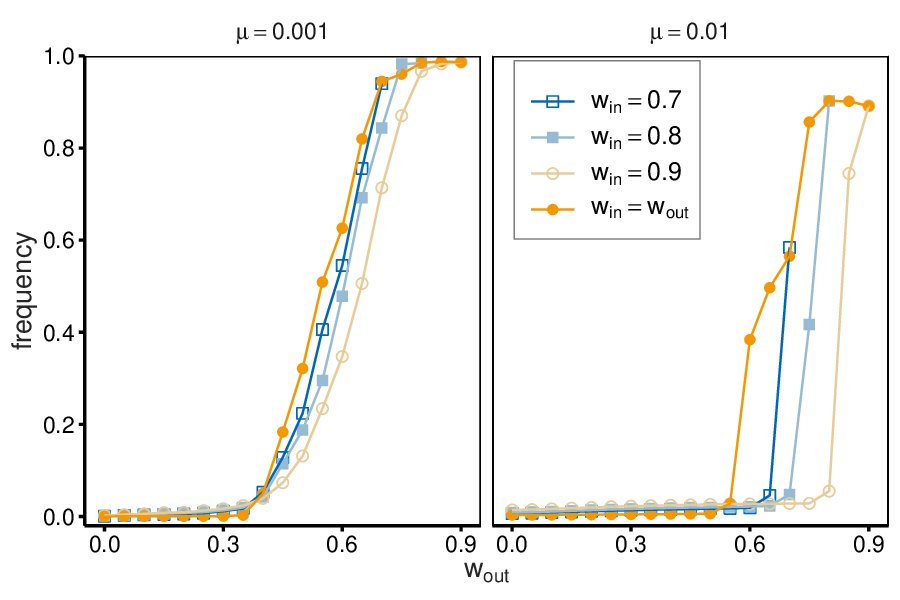}
\caption{\small Frequencies of $XXX\_CCX$ are reported. Large $\win$ ($> \wout$) hinders cooperation with out-group members, even when the values of $\wout$ are sufficiently large to sustain out-group cooperation in the case of $\win = \wout$. Other parameters: $N = 1000, L = 100, b = 3, \sigma = 0$, and $\beta = 1$.}
\label{comp_sigma000}
\end{figure}

Figure~\ref{comp_sigma000} suggests that avoidance of out-group members can be attributed to the availability of relationships with in-group members that are more stable and hence more profitable. This figure reports the frequencies of agents who adopt cooperative strategies toward out-group members ($XXX\_CCX$). It is noted that the frequency of $CCX\_CCX$ does not provide for a fair comparison because large $\win$ enhances in-group cooperation and increases the frequency of $CCX\_CCX$. The evolution of cooperation toward out-group agents requires larger $\wout$ in the case of $\win > \wout$ than in the case of $\win = \wout$. That is, out-group cooperation does not evolve even with a sufficiently large $\wout$ to support cooperation when applied to the entire population. This tendency is more prominent when $\win$ takes larger values; the evolution of out-group cooperation is most inhibited when the value of $\win$ is 0.9. A larger continuation probability implies a favorable environment for cooperation, but cooperation with in-group members slows out-group cooperation in this case; the opportunities for in-group cooperation result in out-group avoidance.

\begin{figure}[tbp]
\centering
\vspace{5mm}
\includegraphics[width = 85mm, trim= 0 0 0 0]{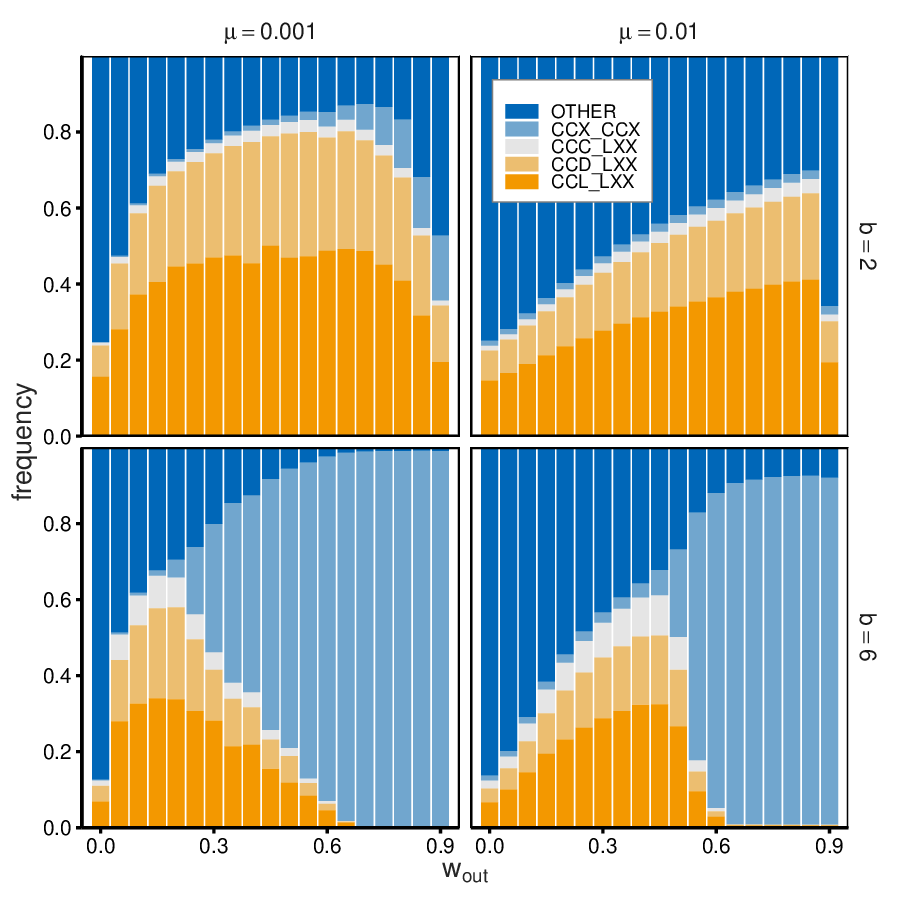}
\caption{\small Frequencies of strategies are reported as a function of $\wout$ when $b = 2$ and $b = 6$. In-group favoritism and out-group avoidance are more frequent when $b = 2$. Other parameters: $N = 1000, L = 100, \win = 0.9, \sigma = 0$, and $\beta = 1$.}
\label{w090_sigma000_b2b6}
\end{figure}

The specific quantitative outcomes can depend on the values of other parameters as fixed in the abovementioned results. Here, we report the role that the benefit of cooperation ($b$) plays in the evolution of out-group avoidance. Figure~\ref{w090_sigma000_b2b6} reports evolutionary outcomes when $b = 2$ and $b = 6$. Cooperation that does not discriminate between the two groups produces large frequencies when the benefit of cooperation is large, whereas in-group favoritism and out-group avoidance flourish where the benefit of cooperation is small. Thus, it follows that squandering the opportunities of cooperative interactions is more harmful where the benefits of cooperation are higher, and out-group avoidance does not pay when this is the setting. More importantly, however, the qualitative patterns of the effects of $\wout$ remain the same. In addition, the suppression of out-group cooperation that is reported in Figure~\ref{comp_sigma000} was observed with these values of the benefit of cooperation (supplementary materials reports this result). 

Finally, we briefly mention the results that arise from different values of the payoff for the leave option ($\sigma$); the detailed simulation outcomes of these are reported in the supplementary materials. This parameter may take a positive value where agents save the cost of participating in games, but its value may also be negative, taking into account the expense on searching partners. Therefore, we examined cases where the value of $\sigma$ is not zero. First, we replicated the main outcomes with the effects of $\wout$ and observed the evolution of out-group avoidance. Second, the values for this parameter have straightforward effects on strategy frequencies. Positive $\sigma$ increased the frequency of out-group avoidance, while negative $\sigma$ decreased it. Thus, positive (negative) payoffs for avoidance increased (decreased) the frequency of avoidance strategies. 

\section*{Conclusion}
This study examined the emergence of in-group favoritism and out-group avoidance in a repeated prisoner's dilemma game, using agent-based modeling. We considered the situation where the continuation probability is higher with in-group members than with out-group members. The simulation results demonstrated that cooperation with in-group members was established, but agents determined to avoid interaction with out-group members even without observing the first move. Comparison with the case of equal continuation probability suggested that a larger continuation probability with in-group members would inhibit out-group cooperation. Cooperation with out-group members did not evolve, even when the continuation probability was sufficiently large for this evolution in a population without effective group distinction. 

The direct implication of the simulations is the potential importance of out-group avoidance. Studies have identified in-group favoritism in social dilemmas \citep{Balliet2014}. These studies have examined situations where interaction is supposed to occur, and participation in games is not optional. This study implies that the different treatment of in- and out-group members in potentially cooperative relationships may take place before the onset of interactions. Research on the evolution of cooperation has elucidated the role that partner choice plays in the establishment of cooperation \citep{Sanchez2015, Barclay2016}. Simultaneously, however, sociological and psychological research has documented homophily and segregation between groups \citep{McPherson2001, Bettencourt2019}. A recent experiment demonstrated the coevolution of cooperation and homophily in dynamic networks \citep{Melamed2020}. Our study advances this research on cooperation and preferential interactions with in-group members through evolutionary agent-based simulations. This simulation suggests that partner selection fosters preferential interactions for in-group members, and this can hinder the realization of cooperative relationships with out-group members. Partner selection can make stable in-group relationships assured by exogenous environment spillover into strategic behavior for social dilemmas. 

The results of simulation might also speak to other observations regarding in-group favoritism. In-group favoritism may not necessarily accompany out-group hate, and discrimination can occur in the absence of aggression toward out-groups \citep{Brewer1999, Mifune2017}. In the context of cooperation for social dilemmas, there is no difference between levels of cooperation with out-group members and those with strangers with an unknown group affiliation \citep{Balliet2014}; in-group favoritism can be observed in the absence of out-group derogation. In simulations, agents prefer interactions with in-group members because those relationships tend to be more stable and profitable. In-group interactions thus serve as an outside option \citep{Baumard2013}. Consequently, in-group favoritism can emerge without hostile intention, where out-group members are simply avoided. More specifically, agents often avoided out-group members, even when the continuation probability was sufficiently high to support cooperation. Here, only interactions with the most profitable candidates, i.e., in-group members, were considered, and all other agents are ignored, which could explain the relative lack of attention paid to the distinction between out-group members and unidentified strangers. Another point that might be related to this study is the question of emotion toward out-groups. The simulation suggested that the main reaction to out-group members was avoidance; therefore, it could be inferred that the emotion triggered by out-group members might be related to avoidance. Evolutionary psychology has demonstrated the function of emotions. For instance, anger can be triggered to cope with threats to group resources and property \citep{Cottrell2005}. Intergroup anxiety serves to avoid or deter threating experiences, which can foster avoidance of contact with out-group members \citep{Stephan2014, Paolini2018}. Our results imply that some types of emotions that prompt avoidance may trigger selective interactions with in-group members in social dilemmas. 

We conclude this paper by discussing the limitations of this research and its possible future extensions in relation to the literature on the evolution of cooperation and in-group favoritism. First, this study considered only direct reciprocity, and indirect reciprocity is not considered. The bounded generalized reciprocity model is an influential theory that explains in-group favoritism through incorporating the framework of indirect reciprocity that stresses behavior that is conditional on the reputation of the partner \citep{Yamagishi2000, Yamagishi2008}. This theory argues that prosocial behavior only enhances fitness when it is targeted at in-group members. Prosocial behavior toward out-group members is not repaid because reputation is beyond the scope of out-group members. We argue that the bounded generalized theory can be potentially combined with the partner selection framework. Reputation can be exploited not only to determine whether to cooperate, but also to choose an interaction partner \citep{Roberts2021}. A large probability of obtaining reputation information can enhance the evolution of cooperation arising through indirect reciprocity \citep{Ghang2015}. The partner selection framework and bounded generalized reciprocity can be combined through consideration of the different probability of obtaining reputation information between groups. A higher propensity to gain reputation information in in-group interactions could make in-group members preferred when seeking interaction partners. 

Several extensions of the model can be considered, beyond considering other hypotheses. This study assumed that a fixed group structure exists exogenously. However, it is possible that the group agents belong to coevolves with strategies \citep{Fu2012}. Group structure can emerge through an evolution of cooperative social networks \citep{Gray2014, Gross2019a}. The endogenous treatment of groups may deepen our understanding of group-dependent strategies in relation with the evolution of group structure. Furthermore, this study assumed the same payoff values for in- and out-group interactions. This assumption implies that avoiding out-group members does not lead to welfare loss in each interaction. However, empirical and theoretical studies have observed that preference for in-group partners often results in the loss of profitable opportunities with out-groups \citep{Greif1989, McConnell2018, Takesue2020}. Differences in payoffs can contribute to developing an understanding about the welfare implications of out-group avoidance. Finally, our study did not consider out-group attack, which has become an increasingly important research topic \citep{DeDreu2020}. Research subjects include causes of intergroup conflict, such as carrying-capacity stress \citep{DeDreu2022} and emotions that trigger aggression \citep{Matsumoto2015}. While this point goes beyond the scope of this research, future studies should ultimately address conditions that could explain the presence and absence of enmity against out-groups. We believe that research on the role that partner selection plays in in-group favoritism promises fruitful results. 

\bibliographystyle{unsrt}
\bibliography{ingroup_partner}



\end{document}


\maketitle

\begin{figure*}[tbp]
\centering
\vspace{5mm}
\includegraphics[width = 140mm, trim= 0 0 0 0]{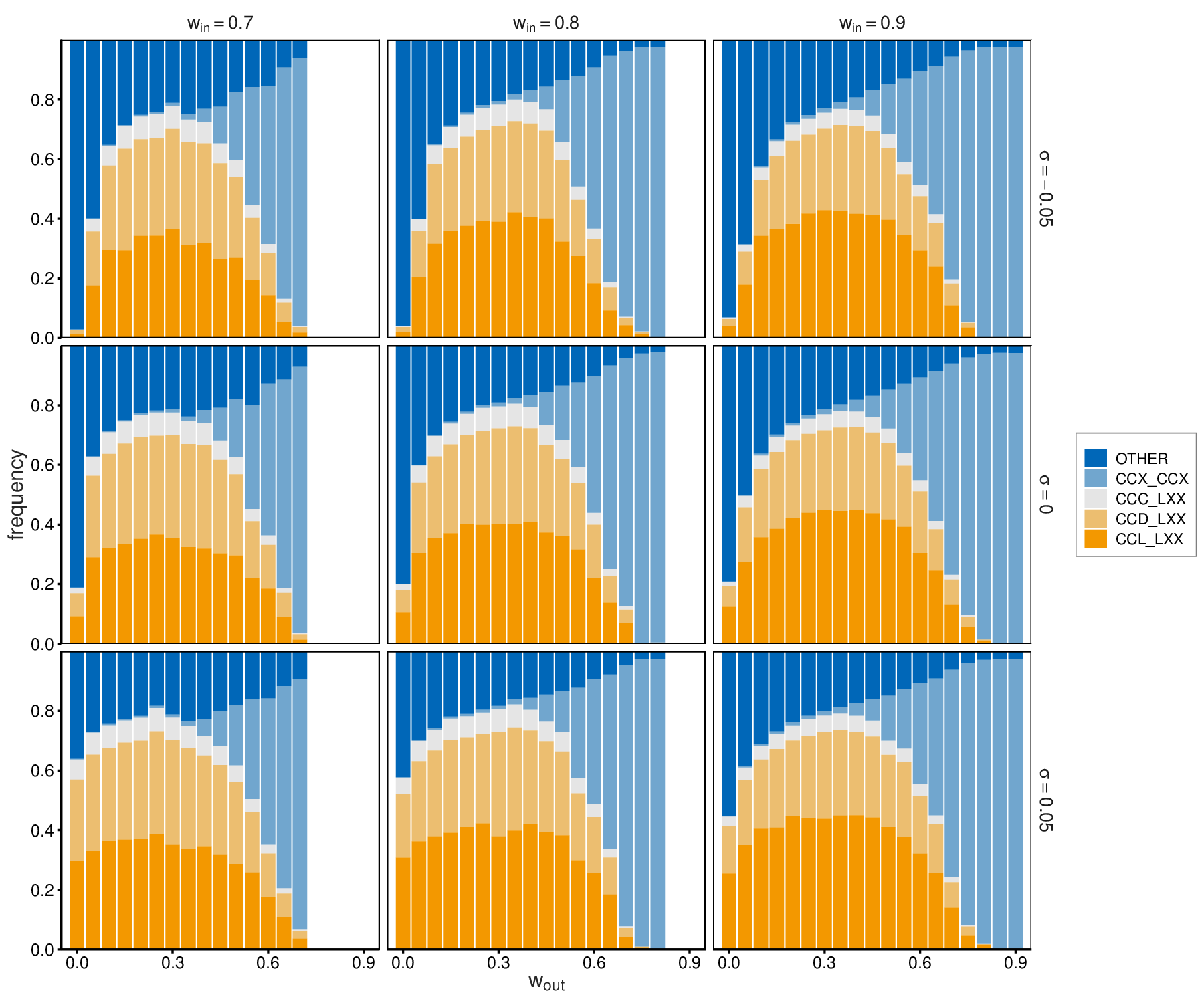}
\caption{\small Frequencies of strategies when $\mu = 0.001$ are reported. Other parameters: $N = 1000, L = 100, b = 3$, and $\beta = 1$.}
\label{w_w_sigma_mu001}
\end{figure*}

\begin{figure*}[tbp]
\centering
\vspace{5mm}
\includegraphics[width = 140mm, trim= 0 0 0 0]{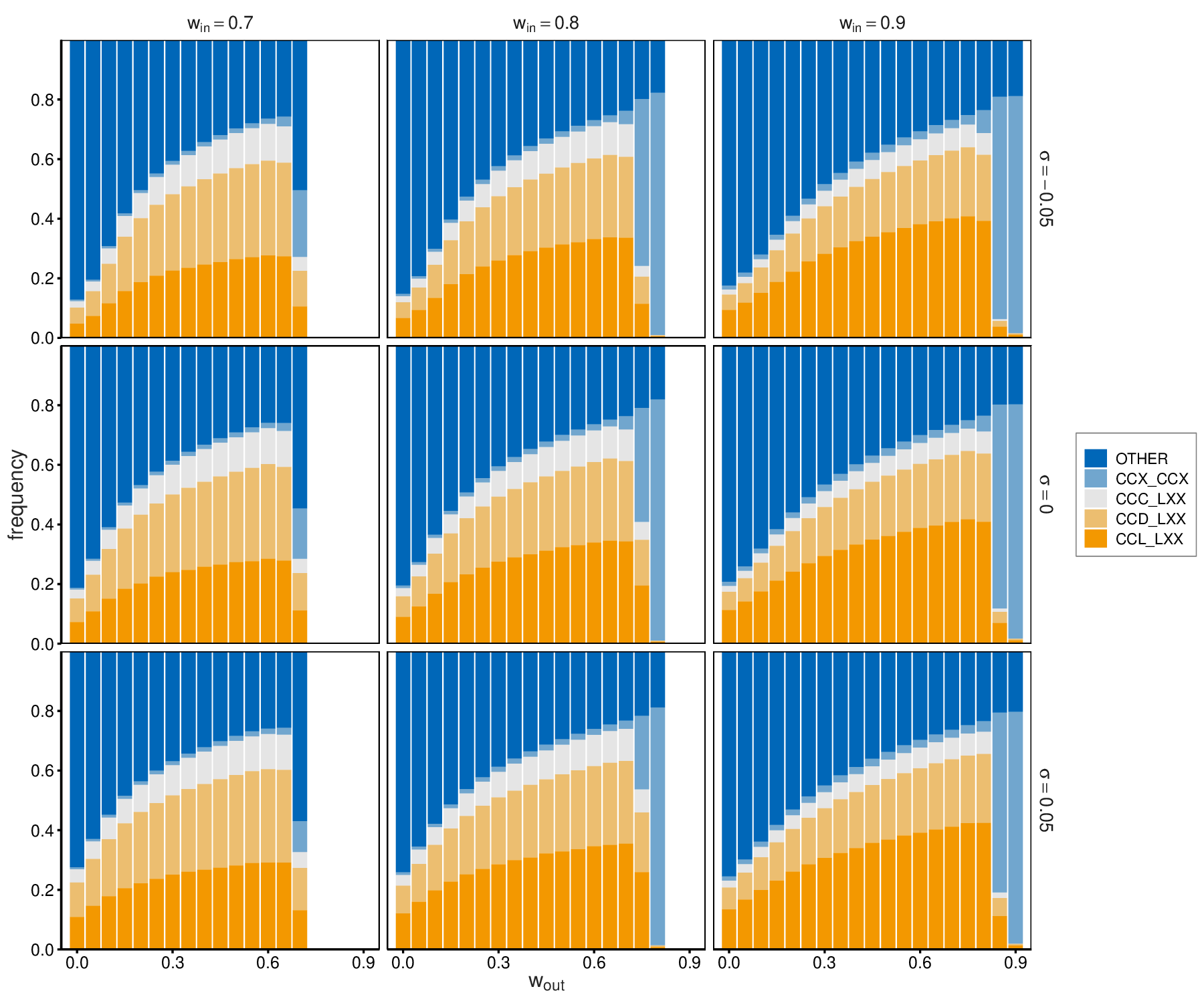}
\caption{\small Frequencies of strategies when $\mu = 0.01$ are reported. Other parameters: $N = 1000, L = 100, b = 3$, and $\beta = 1$.}
\label{w_w_sigma_mu01}
\end{figure*}

In this supplementary material, we report the results of simulation, some of which are not reported in the main text. First, we report the results where the values of $\sigma$ are not zero. From this setting, agents acquire positive or negative payoffs through selection of the leave option. Positive values may correspond to the saved participation cost of games, whereas negative values may correspond to the search cost for a new partner. Figure~\ref{w_w_sigma_mu001} reports the results, along with those of cases of $\sigma = 0$ when the mutation rate is 0.001. The values for $\win$ are 0.7, 0.8, and 0.9, respectively. The figure shows that basic patterns are not influenced by the values for $\sigma$ (and $\win$). Cooperation toward in-groups and avoidance of out-groups are observed with medium values of $\wout$. The effect of $\sigma$ is straightforward; positive $\sigma$ values support the evolution of strategies specifying the leave option toward out-groups, whereas negative $\sigma$ hinders it. This pattern is clearly discernible for small values of $\wout$. Figure ~\ref{w_w_sigma_mu01} indicates that the same results appear for a mutation rate of 0.01, and similar patterns are replicated.

\begin{figure*}[tbp]
\centering
\vspace{5mm}
\includegraphics[width = 140mm, trim= 0 0 0 0]{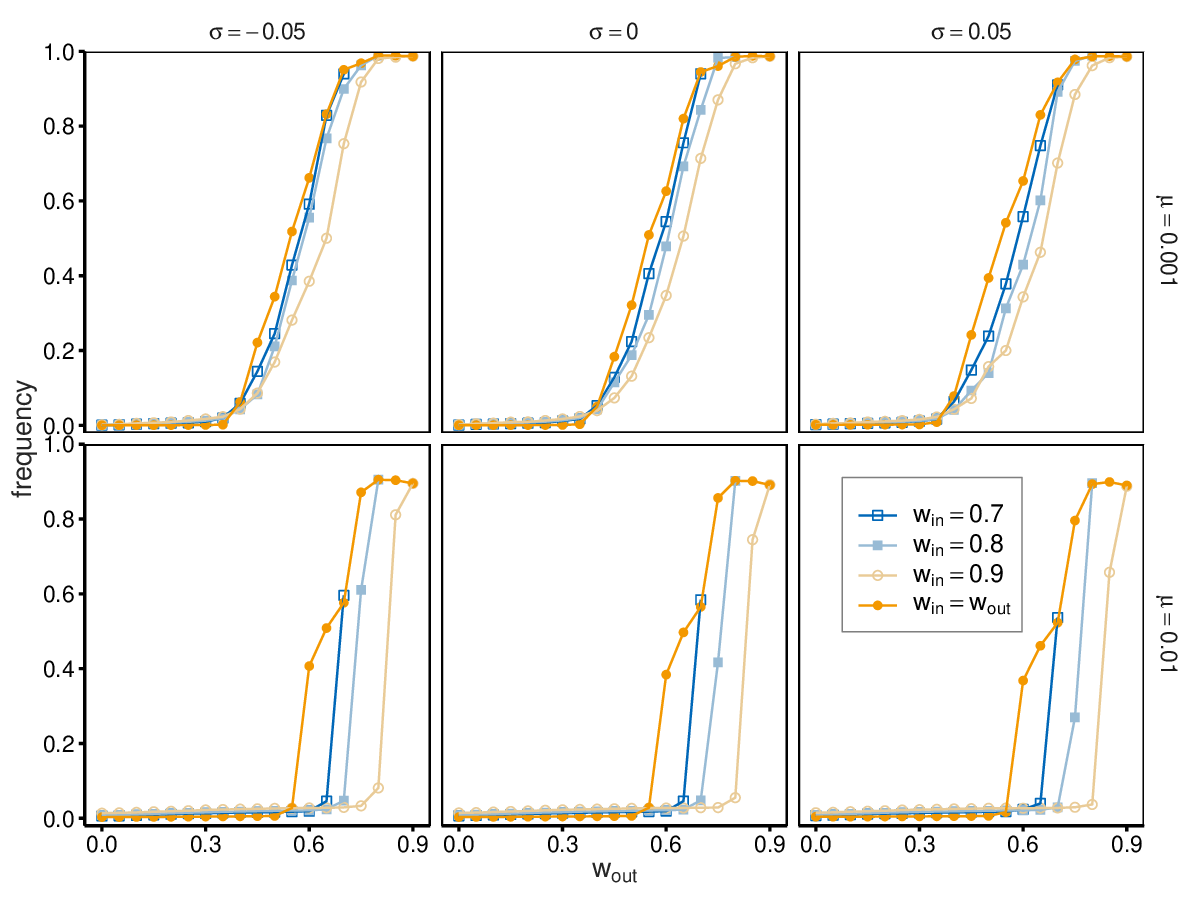}
\caption{\small Cases with fixed values of $\win$ and those under $\win = \wout$ are compared. Other parameters: $N = 1000, L = 100, b = 3$, and $\beta = 1$}.
\label{comp_w_w_sigma}
\end{figure*}

We examine the frequencies for $XXX\_CCX$ and compare these with those for the condition that $\win = \wout$. Figure~\ref{comp_w_w_sigma} indicates that cooperation toward out-group members evolves through smaller values of $\wout$ than is the case for $\win > \wout$. The main text reports this pattern when $\sigma = 0$, and this pattern is replicated where $\sigma \neq 0$. These results indicate that increased stable interaction opportunities with in-group members reduce cooperation with out-group members. 

\begin{figure*}[tbp]
\centering
\vspace{5mm}
\includegraphics[width = 90mm, trim= 0 0 0 0]{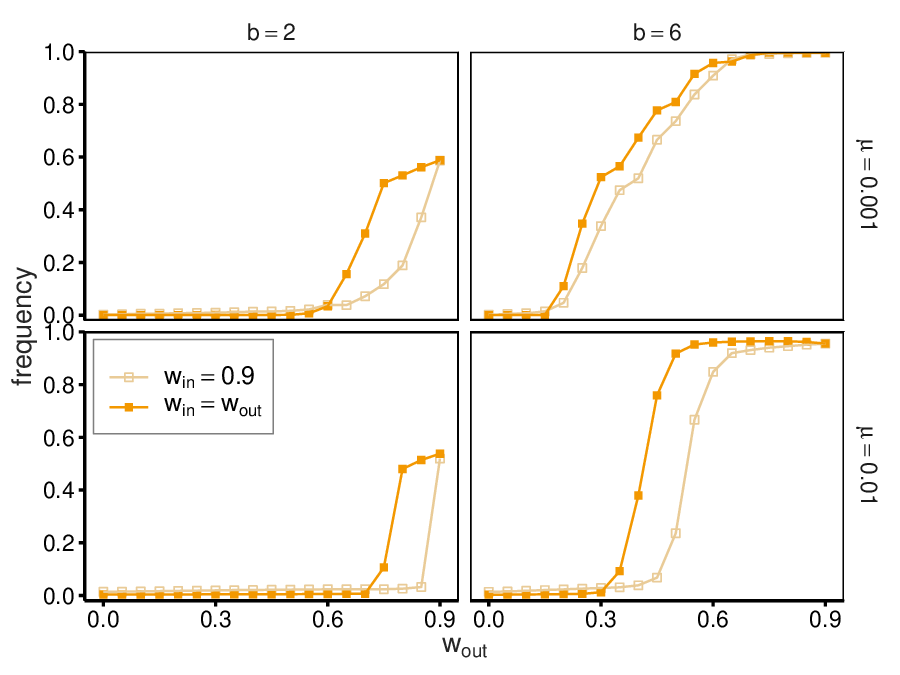}
\caption{\small Cases with fixed values of $\win$ and those under $\win = \wout$ are compared. Other parameters: $N = 1000, L = 100, \sigma = 0$, and $\beta = 1$.}
\label{comp_w_b_mu}
\end{figure*}

In addition, we observe this pattern with different values for the benefit of cooperation ($b$). The main text reports the frequency of strategies in the cases of $b = 2$ and $b = 6$, and in this supplement, we compare these results with the cases of $\win = \wout$. Figure~\ref{comp_w_b_mu} replicates the main observation that the evolution of cooperation toward out-group members requires larger $\wout$ for $\win > \wout$ than it does for $\win = \wout$. 

